\documentclass[11pt]{article}
\usepackage{epsf}
\begin{document}
\newcommand{\prl}[1]{Phys. Rev. Lett. {\bf #1}}
\newcommand{\prev}[1]{Phys. Rev. {\bf #1}}
\newcommand{\prd}[1]{Phys. Rev. D {\bf #1}}
\newcommand{\zs}[1]{Z. Phys. {\bf #1}}
\newcommand{\ncim}[1]{Nuovo Cim. {\bf #1}}
\newcommand{\plet}[1]{Phys. Lett. {\bf #1}}
\newcommand{\prep}[1]{Phys. Rep. {\bf #1}}
\newcommand{\rmp}[1]{Rev. Mod. Phys. {\bf #1}}
\newcommand{\nphy}[1]{Nucl. Phys. {\bf #1}}
\newcommand{\nim}[1]{Nucl. Instrumen. Meth. {\bf #1}}
\newcommand{\et}{{\rm E}_{\scriptscriptstyle\rm T}}
\newcommand{\met}{\mbox{$\protect \raisebox{.3ex}{$\not$}\et$}}
\begin{flushright}
FERMILAB-CONF-97/166-E\\
\today
\end{flushright}
\vspace*{3.0cm}
\begin{center}
\begin{large}
\textbf{TOP QUARK PRODUCTION AND DECAY
        AT THE TEVATRON}\footnote{To appear in the proceedings of
the XXXIInd Rencontres de Moriond, Electroweak Interactions and 
Unified Theories, Les Arcs, Savoie, France, March 15-22, 1997.} \\
\end{large}
\vspace*{22pt}
                David W. Gerdes \\
                (representing the CDF Collaboration) \\
\vspace*{15pt}
\textit{Department of Physics and Astronomy \\ The Johns Hopkins University\\
 3400 North Charles Street, Baltimore, Maryland 21218 } \\
\end{center}

\begin{figure*}[h]
\begin{center}
\leavevmode
\epsfysize=2.9in
\epsffile{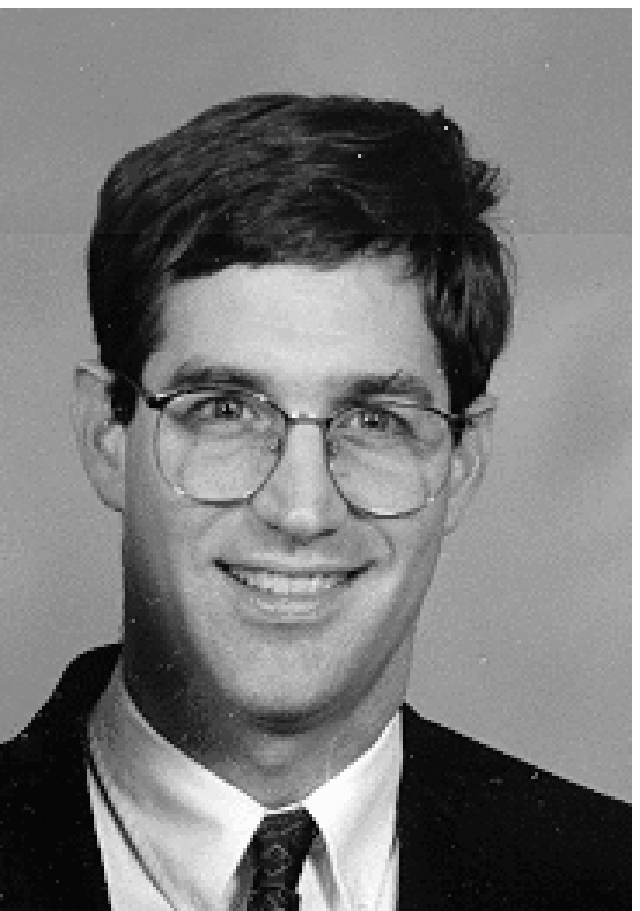}
\end{center}
\end{figure*}
\vspace*{0.5cm}
\textbf{Abstract:} Recent measurements of the top quark production cross
section and decay properties by the CDF and D0 experiments are described. 
The cross section
has been measured in dilepton, lepton plus jets, and all-hadronic final
states, and a measurement of $BR(t\rightarrow Wb)/BR(t\rightarrow Wq)$,
where $q$ is any quark,
has been performed. The results, though statistics-limited, are consistent 
with each other and with theoretical predictions.
\newpage
\section{Overview of Top Quark Production}
\baselineskip 15pt
Since the observation of the top quark in 1995 by the CDF\cite{cdf-obs}
and D0\cite{d0-obs} collaborations, the Tevatron experiments have moved
rapidly into a program of detailed studies of the top quark. In this paper
I describe recent measurements of the top quark production cross 
section\cite{d0-xsec,cdf-xsec} by CDF and D0 
using a number of final state topologies. I also describe a recent CDF
measurement of $BR(t\rightarrow Wb)/BR(t\rightarrow Wq)$, where $q$ is
any quark. Recent measurements of the top quark mass are described in 
Ref.~\cite{mtop-refs}.

At the Tevatron energy of $\sqrt{s}=1.8$~TeV, most top quarks are produced
in pairs via the annihilation processes $q\bar{q}\rightarrow t\bar{t}$ (90\%)
and $gg\rightarrow t\bar{t}$ (10\%). Top quarks can also be produced singly
by electroweak processes such as $W$-gluon fusion. While no signal has
been isolated for the single-top process yet, it is included as a 
``background'' when studying $t\bar{t}$ production. For the remainder of
this paper we consider only the pair production of top quarks.

In the Standard Model each top quark decays nearly 100\% of the time into
$Wb$. Each $W$ in turn can decay into a charged lepton plus neutrino, 
with a branching
ratio of 1/9 to each lepton family, or into a $q\bar{q}^{\prime}$ pair
(``jets''), with a branching ratio of 2/3. Top quark candidate events
are characterized by the decay modes of the two $W$'s. Most analyses done
by the Tevatron experiments focus on final states containing at least
one $W$ decay to $e\nu$ or $\mu\nu$:
\begin{itemize}
   \item \textbf{Dilepton} final states (5\% of $t\bar{t}$ decays) contain
   two isolated, high-$P_T$ leptons ($e^+e^-$, $\mu^+\mu^-$, or 
   $e^{\pm}\mu^{\mp}$), significant missing transverse energy ($\met$) from
   the undetected neutrinos, and two jets from $b$ quarks.
   \item \textbf{Lepton + jets} final states (30\% of $t\bar{t}$ decays)
   contain one isolated, high-$P_T$ electron or muon, significant $\met$,
   and typically three or more jets, two of which are from $b$'s.
\end{itemize}
While this paper will emphasize these two final state topologies, top 
decays to all-hadronic final states have been observed by 
CDF\cite{all-had-prl}, and a handful of suggestive events have also been 
observed in the $\tau$ dilepton channel\cite{tau-dilep}.

\section{$t\bar{t}$ Production Cross Section}

The top quark production cross section, $\sigma_{t\bar{t}}$, is of interest
for several reasons. First, it is a test of QCD calculations\cite{xsec-berger,
xsec-catani,xsec-laenen}. Second, departures from the theoretical expectation
could indicate new physics, such as production through a high-mass intermediate
state or decays to final states other than $Wb$. By measuring 
$\sigma_{t\bar{t}}$ in as many channels as possible, we hope to gain a
consistent picture of top as a Standard Model object or to identify the
places where the theory may be in error.
Finally, $\sigma_{t\bar{t}}$
is an important ``engineering number'' for estimating top yields in future
experiments at the Tevatron and LHC\cite{TEV2000}.
The $t\bar{t}$ production cross section has been measured in the dilepton,
lepton + jets, and all-hadronic final states using the full Run I datasets
with integrated luminosities of approximately 110 pb$^{-1}$.

\subsection{Dilepton Analysis}

Dilepton events result from the process $t\bar{t}\rightarrow WbW\bar{b}
\rightarrow \ell^+\nu b \ell^-\bar{\nu}\bar{b}$.
The CDF dilepton analysis begins with a single inclusive lepton sample
that also forms the starting point for the lepton plus jets analysis. Events
in this sample contain an isolated $e$ or $\mu$ with $P_T>20$~GeV and
pseudorapidity $|\eta|<1$. A second, opposite-charge $e$ or $\mu$ is then 
required with 
$P_T> 20$~GeV. The second lepton may satisfy looser quality cuts. Because
top dilepton events contain two $b$ jets, two jets are required with
observed transverse energy\footnote{\baselineskip 11pt 
Observed jet energies differ from the true parton energy because of 
both instrumental and physics effects. Instrumental effects include 
detector nonlinearity and cracks. In addition, fragmentation effects can cause
energy to be deposited outside the jet clustering cone, and unrelated
energy from multiple interactions or the underlying event can be deposited
inside the clustering cone. CDF cuts on observed jet energies, while D0 
applies a correction factor. The ratio between corrected and observed jet
energies at CDF is rapidity- and $E_T$-dependent but averages approximately 
1.4.}
$E_T > 10$~GeV and $|\eta|<2.0$. At least 25~GeV of
$\met$ is required. If $\met < 50$ GeV, the angle between the $\met$ 
vector and the nearest lepton or jet is required to be at least 15$^{\circ}$.
This cut reduces backgrounds from $Z\rightarrow\tau\tau$ and mismeasured
jets. Finally, $ee$ and $\mu\mu$ events with a dilepton invariant mass
in the $Z$ mass window between 75 and 105~GeV are removed, as are $ll\gamma$
events with a three-body invariant mass consistent with a radiative $Z$ decay.

Nine candidates remain in the final CDF dilepton sample: 
one $ee$, one $\mu\mu$,
and seven $e\mu$. Four of the nine events are $b$-tagged using the algorithms
described below.
The relative numbers of events are consistent with the expectations from
$t\bar{t}$ Monte Carlo ($M_{top}=175$~GeV), which predicts relative 
acceptances of 15\%, 27\%, and 58\% in the $ee$, $\mu\mu$, and $e\mu$ 
channels respectively. The background is calculated to be
2.1$\pm$0.4 events, and consists of lepton pairs from the Drell-Yan process,
$Z\rightarrow \tau\tau$, $W$ pair production, and fakes. 

The D0 dilepton analysis also makes use of $ee$, $\mu\mu$, and $e\mu$ final
states, with cuts similar to those described above. Electrons are searched
for in the range $|\eta|<2.5$, and muons in the range $|\eta|<1.7$. The
lepton $P_T$ threshold is 15~GeV for the $\mu\mu$ and $e\mu$ analyses and
20~GeV for the $ee$ analysis. At least two jets with corrected 
$E_T > 20$~GeV and
$|\eta|<2.5$ are required. In addition, since top events tend to have rather
energetic jets, a cut is placed on the scalar summed transverse energy of
the jets with $E_T > 15$~GeV plus the leading electron, if present. 
One $ee$, one $\mu\mu$, and three $e\mu$ events are observed, with a 
background of 1.4$\pm$0.4 events.

\subsection{Lepton Plus Jets Analysis}

Lepton plus jets events arise from $t\bar{t}\rightarrow WbW\bar{b}
\rightarrow \ell\nu b q\bar{q}^{\prime}\bar{b}$. Four jets are therefore expected
in the final state, two from $b$'s and two from the hadronic $W$ decay.
However, jets may be merged or lost due to detector effects, and additional
jets may be produced from gluon radiation. The CDF and D0 analyses therefore
begin by requiring an isolated lepton with $P_T>20$~GeV, significant
$\met$, and at least three jets. There remains a significant QCD background
from $W$ plus multijet production, which can be reduced to acceptable
levels through kinematic cuts or $b$-tagging.

The D0 analysis takes two complementary approaches. The ``$\ell+$jets/$\mu$''
analysis seeks to tag a $b$ jet by identifying a muon from $b\rightarrow
\mu X$ in the vicinity of a jet. At least three jets are required with
corrected $E_T>20$ and $|\eta|<2$. The tagged muon is required to have $P_T > 4$~GeV
and to be within $\Delta R = \sqrt{\Delta\eta^2 + \Delta\phi^2}\le 0.5$ of
a jet. Loose cuts are placed on the summed transverse energy of the jets,
$H_T>110$~GeV, and on the event's aplanarity, ${\cal A}>0.040$. Eleven events
are observed on a background of $2.4\pm 0.5$ events. The dominant backgrounds 
are fake leptons, which are estimated from control samples in the data, and
QCD $W$ plus multijet production, which is modelled using the 
\textsc{vecbos}\cite{Vecbos} event generator interfaced to the 
\textsc{herwig}\cite{Herwig} parton shower model and passed through a full 
detector simulation. The background $\mu$-tag rate, which includes both fake
tags as well as real heavy flavor in the background, is estimated from
multijet data.

\begin{figure}[htbp]
\begin{center}
\leavevmode
\epsfysize=2.6in
\epsffile{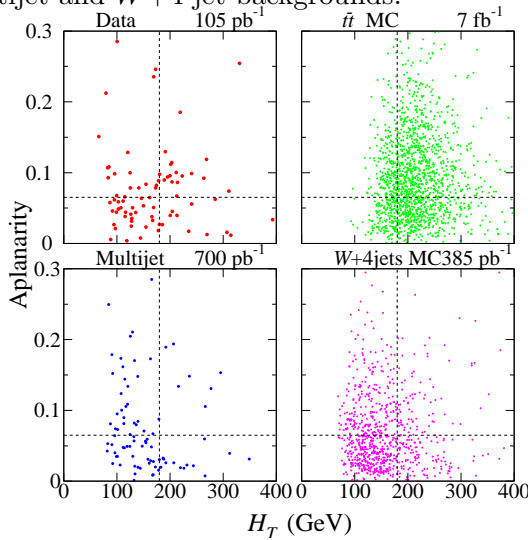}
\caption{\baselineskip 11pt
Distributions of ${\cal A}$ vs. $H_T$ in the D0 $\ell+$jets 
analysis. Clockwise from upper left: $\ell+$jets data; $t\bar{t}$ Monte 
Carlo ($M_{top} = 175$~GeV); $W+$ 4-jet Monte Carlo; multijet background.
Events above and to the right of the dashed lines pass the kinematic 
selection.}
\label{fig:D0ljet}
\end{center}
\end{figure}

The second D0 approach to the lepton plus jets channel makes use of 
kinematic information to distinguish $t\bar{t}$ events from the
$W$ plus multijet background. Top events generically have more energetic jets 
and are more spherical than the background, so the kinematic variables
$H_T$ and ${\cal A}$, defined above, are expected to have discriminating
ability. A Monte Carlo optimization procedure is used to select the cuts 
on $H_T$ and ${\cal A}$ that minimize the expected cross section uncertainty.
Events are required to have at least four jets with corrected $E_T>15$~GeV and
$|\eta|<2$, to have $\met>25(20)$~GeV in the electron (muon) channel,
and to satisfy $H_T > 180$~GeV, ${\cal A}>0.065$. In addition, the scalar
sum of the lepton transverse energy and the $\met$ is required to exceed
60~GeV, and the leptonically-decaying $W$ is required to be loosely
central, satisfying $|\eta_W|<2$. Events that pass the $\ell+$jets/$\mu$
selection are excluded from this analysis. These kinematic cuts identify 19 
events, with a background of 8.7$\pm$1.7 events. Figure~\ref{fig:D0ljet}
shows the distribution in the $H_T$-${\cal A}$ plane of the data, $t\bar{t}$
signal Monte Carlo, and multijet and $W+$4-jet backgrounds.

The CDF lepton + jets analysis begins with inclusive electron and muon
samples, where the primary lepton is required to have $P_T > 20$~GeV,
$|\eta|<1$, and to be isolated. At least 20~GeV of $\met$ is required.
At least three jets with observed 
$E_T>15$~GeV and $|\eta|<2$ are then required.
A total of 325 events pass these cuts, with a signal-to-background ratio
of about 1:4.
CDF uses two $b$-tagging techniques to reduce the $W$ plus multijet background
in this sample: soft lepton (SLT) tagging, and secondary vertex (SVX) tagging.
The SLT tag requires an electron or muon with $P_T>2$~GeV in the vicinity
of one of the jets. This technique has an efficiency of 20$\pm$2\% for
$t\bar{t}$ events that pass the initial selection, and has a typical fake
rate per jet of 2\%. 
The SVX technique uses precision tracking information
from the silicon vertex detector\cite{cdf-svx} to reconstruct secondary
vertices from $b$ decays. The efficiency of this technique is 41$\pm$4\%,
with a typical fake rate per jet of 0.5\%. Because of its high efficiency
and low background, SVX-tagging is CDF's primary $b$-tagging technique.

\begin{figure}[htbp]
\begin{center}
\epsfysize=2.6in
\leavevmode
\epsffile[0 100 567 620]{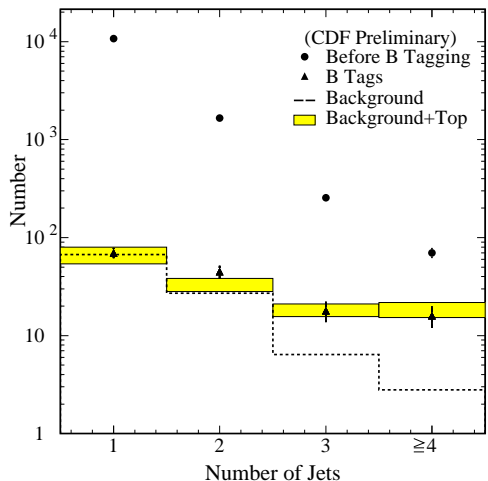}
\caption{\baselineskip 11pt
Number of SVX-tagged jets as a function of jet multiplicity.
The $t\bar{t}$ signal region with $N_{jet}\ge 3$ shows a large excess of
tags.}
\label{fig:cdf-ljet}
\end{center}
\end{figure}

After SVX-tagging, 34 events are identified on a background of 8.0$\pm$1.4.
Eight of the events have two SVX-tagged jets. The background is dominated
by real heavy flavor ($Wb\bar{b}$, $Wc\bar{c}$). Figure~\ref{fig:cdf-ljet}
shows the number of SVX-tagged jets as a function of jet multiplicity.
The SLT technique
identifies 40 events on a background, dominated by fakes, of 24$\pm$3.5.
\subsection{Other Channels}

CDF has observed a $t\bar{t}$ signal and measured the cross section in the 
all-hadronic channel using a combination of kinematic cuts and 
SVX-tagging\cite{all-had-prl}. CDF has also reported\cite{tau-dilep} a 
modest excess of events in dilepton final states containing a $\tau$ 
candidate. D0 has increased their acceptance for $t\bar{t}$ production
by including events with an isolated electron with $E_T>20$~GeV and
$|\eta|<1.1$, at least 2 jets with corrected $E_T>30$~GeV, large missing
energy, $\met>50$~GeV, and high transverse mass, $M_T(l$-$\met)>115$~GeV. 
Events that pass the standard dilepton or lepton
plus jets cuts are excluded. This selection provides sensitivity to
$\tau$ decays and regains some dilepton and lepton
plus jets events that fail the standard kinematic selection, for example
because a lepton or jet was lost or mismeasured. Four events pass this
``$e\nu$'' selection, with a background of 1.2$\pm$0.4 events.

\subsection{Cross Section Results}

The $t\bar{t}$ acceptance is evaluated using the \textsc{herwig}
event generator together with a detector simulation. Lepton
identification and $b$-tagging efficiencies are corrected, where necessary,
using values measured in the data. The acceptance is a slowly-rising function
of $M_{top}$. To quote a cross section, CDF uses a top mass of 175~GeV,
while D0 uses 173.3~GeV. Backgrounds are rescaled 
to account for the $t\bar{t}$ component of the data.
The results from the various channels are
shown in Table~\ref{tab:xsec_cdf}.

\begin{table}[htbp]
\begin{center}
\leavevmode
\caption{\baselineskip 11pt 
Top quark production cross section results from the Tevatron
experiments. Acceptances and cross sections are evaluated at 
a top mass of 175 GeV for CDF and 173.3 GeV for D0.}
\begin{tabular}{cccccc}
\hline\hline
Channel & Acceptance (\%) & $\int{\cal L}\,dt$ (pb$^{-1}$)& Background &$N_{obs}$ & $\sigma_{t\bar{t}}$ (pb) \\ \hline
Dilepton (CDF) & 0.74$\pm$0.08 & 109 & 2.1$\pm$0.4 & 9 & $8.5^{+4.4}_{-3.4}$ \\
Dilepton (D0)  & 0.64$\pm$0.11 & 125,105,108($ee,\mu\mu,e\mu$) & 
  1.4$\pm$0.4 & 5 &  \\
$e\nu$ (D0) & 0.28$\pm$0.08 & 108 & 1.2$\pm$0.4 & 4 & 6.3$\pm$3.3$^*$   \\
$\ell+$jets/SVX (CDF)& 3.5$\pm$0.7 & 109 & 8.0$\pm$1.4 & 34 & $6.8^{+2.3}_{-1.8}$ \\
$\ell+$jets/SLT (CDF) & 1.7$\pm$0.3 & 109 & 24.3$\pm$3.5 & 40 & $8.0^{+4.4}_{-3.6}$ \\
$\ell+$jets/$\mu$ (D0) &0.98$\pm$0.15 & 107 & 2.4$\pm$0.5 & 11 & 8.2$\pm$3.5 \\
$\ell+$jets/kin (D0) & 2.32$\pm$0.45 & 110 & 8.7$\pm$1.7 & 19 & 4.1$\pm$2.0\\
All-hadronic (CDF)      & 4.7$\pm$1.6 & 109 & 137$\pm$11 & 192  & $10.7^{+7.6}_{-4.4}$ \\
\hline
\end{tabular}
\label{tab:xsec_cdf}
\begin{flushright}
\vspace*{-0.5cm}
$^*$D0 Dilepton + $e\nu$ combined. 
\end{flushright}
\end{center}
\end{table}
The results are consistent among the different channels, though in some cases
the uncertainties are large. Combining the dilepton and $\ell+$jets 
channels, D0 obtains
\begin{displaymath}
     \sigma_{t\bar{t}}(M_{top}=173.3) = 5.5\pm1.8~{\rm pb~(D0).}
\end{displaymath}
Combining the dilepton and $\ell+$jets channels, CDF obtains
\begin{displaymath}
     \sigma_{t\bar{t}}(M_{top}=175) = 7.5^{+1.9}_{-1.6}~{\rm pb~(CDF).}
\end{displaymath}
For comparison, a recent calculation by Catani 
\textit{et al.}\cite{xsec-catani} gives $\sigma_{t\bar{t}}(175)= 4.75^{+0.73}_{-0.62}$~pb, while 
Berger and Contapaganos\cite{xsec-berger} obtain 
$\sigma_{t\bar{t}}(175)=5.52^{+0.07}_{-0.42}$~pb. 

\section{Measurement of $BR(t\rightarrow Wb)/BR(t\rightarrow Wq)$}

The ratio of branching ratios $B = BR(t\rightarrow Wb)/BR(t\rightarrow Wq)$,
where $q$ is any quark, is predicted to be nearly one in the Standard Model.
CDF has measured $B$ from the ratios double $b$-tagged, single $b$-tagged,
and un-tagged dilepton and lepton plus jets events. Using the known efficiency
for tagging a $b$ jet, which is measured in $b$-enriched control samples,
and a Monte Carlo model of the $b$ jet acceptance in top events, $B$ can 
be extracted from a likelihood fit.

The dilepton sample used for this analysis is the same one used 
in the cross section measurement described above. The $\ell+$jets sample
begins with the W+$\ge$3-jet sample (325 events) used in the cross-section 
measurement. The jets are required to have $E_T>15$~GeV and $|\eta|<2$. 
Then a fourth jet with observed $E_T>8$~GeV and $|\eta|<2.4$ is required, giving
a sample of 163 events. The four-jet requirement facilitates jet-parton
association when comparing data to $t\bar{t}$ Monte Carlo, 
and reduces the $W$ plus
multijet background. Events are classified into four non-overlapping
subsamples: no tags, SLT tags but no SVX tags, single SVX tags, and
double SVX tags. The number of background tags in the various subsamples
is determined through an iterative rescaling as in the cross section
measurement. The observed number of events, the backgrounds and the
$b$-tagging efficiencies per jet are combined into a likelihood fit for
$B$, resulting in
\begin{eqnarray*}
          B & = & 0.99\pm 0.29~{\rm (stat.+syst.)} \\
            & > & 0.58(0.64)~{\rm at~95(90)\%~C.L.}
\end{eqnarray*}
It is important to note that there is some model-dependence in this analysis.
The Monte Carlo model used to calculate the $t\bar{t}$ acceptance assumes
that all top decays are to $Wq$, i.e. that there are no top decays to non-$W$
final states. For example, if the decay $t\rightarrow H^+b$ occurred
with a sizable branching fraction, and if $M_{H^+}\approx M_W$, it would
result in $b$-tagged events that are kinematically identical to ordinary
$\ell+$jets events. Such events would destroy the interpretation of the
measured tag ratios as a measurement of $B$. However, a large branching ratio
of top to non-$W$ final states would also result in fewer than expected
dilepton events and, therefore, a lower than expected cross section measurement
in this channel. Work to combine all available information into limits on
nonstandard decays is in progress.

This result can be converted into a lower limit on $|V_{tb}|$, albeit with
additional assumptions. Assuming a three-generation unitary CKM matrix,
this measurement gives $|V_{tb}|>0.76$ at the 95\% C.L. However, in this
case $|V_{tb}|$ is much better determined from unitarity and independent 
measurements of the other CKM parameters---in fact it is the 
\textit{best-known} CKM matrix element. In the case of four quark 
generations, there are additional CKM angles and phases, and it is not
possible to fix $|V_{tb}|$ by a single measurement without making further
assumptions. Even then, only weak constraints on $|V_{tb}|$ can
be obtained.

\section{Conclusions} 

The top quark production cross section has been measured in a number of 
final states both with and without $b$-tagging, and a measurement of
the ratio of branching ratios $BR(t\rightarrow Wb)/BR(t\rightarrow Wq)$
has been performed.
Taken as a whole, the data are starting to paint a consistent picture of 
the top quark as a Standard Model object. However, the measurements
are statistics-limited, and exotic production mechanisms and large
nonstandard branching ratios can not yet be ruled out\cite{kane-gluino}.

The Tevatron experiments are currently undergoing major upgrades
in preparation for running with the Main Injector in late 1999. In addition
to the expected factor of 20 increase in integrated luminosity, and
the 40\% increase in the top production cross section that will come
from raising the Tevatron energy from 1.8 to 2.0 TeV, the
detector upgrades will result in significant improvements in $b$-tagging
and lepton identification. All told, a factor of 50 increase in the size
of the top sample appears feasible in the first few years of Main Injector
running, with correspondingly bright prospects for precision top physics
at the Tevatron.

\section{Acknowledgements}
I would like to thank the organizers for a very enjoyable ``Rencontre.''
Herb Greenlee and Rich Partridge provided me with details of the D0 analyses. 
This work is supported by NSF grant PHY-9604893 and by a DOE Outstanding 
Junior Investigator award.


\begin{thebibliography}{99}
\baselineskip 11pt
\bibitem{cdf-obs} F.~Abe \textit{et al.}, \prl{74}, 2626 (1995).
\bibitem{d0-obs} S.~Abachi \textit{et al.}, \prl{74}, 2632 (1995).
\bibitem{d0-xsec} S.~Abachi \textit{et al.}, hep-ex 9704015, submitted to
Phys. Rev. Lett.
\bibitem{cdf-xsec} F.~Abe \textit{et al.}, in preparation.
\bibitem{mtop-refs} R.~Raja, these proceedings; S. Abachi \textit{et al.},
   hep-ex/9703008; D.~Gerdes, hep-ex/9609013.
\bibitem{all-had-prl} F.~Abe \textit{et al.}, FERMILAB-PUB-97/075-E, 
  to appear in Phys. Rev. Lett.
\bibitem{tau-dilep} F.~Abe \textit{et al.}, hep-ex 9704007, submitted to
 Phys. Rev. Lett.
\bibitem{xsec-berger} E.~L.~Berger and H.~Contopaganos, \prd{54}, 3085 (1996).
\bibitem{xsec-catani} S. Catani, M.~L.~Mangano, P.~Nason, and L.~Trentadue,
        Phys. Lett. {\bf B378}, 329 (1996).
\bibitem{xsec-laenen} E.~Laenen, J.~Smith, and W.~L.~Van Neerven, Nucl.
   Phys. {\bf B369}, 543 (1992); E.~Laenen, J.~Smith, and W.~L.~Van Neerven,
   Phys. Lett. {\bf B321}, 254 (1994).
\bibitem{TEV2000} See, for example, ``Future Electroweak Physics at the 
    Fermilab Tevatron:
    Report of the TEV2000 Study Group,'' D.~Amidei and R.~Brock eds.,
    Fermilab-Pub-96/082 (1996).
\bibitem{Vecbos} F.~A.~Berends, H.~Kuijf, B.~Tausk, and W.~T.~Geile,
 Nucl. Phys. \textbf{B357}, 32 (1991).
\bibitem{Herwig} G.~Marchesini \textit{et al.}, Comp. Phys. Comm. 
  \textbf{97}, 465 (1992).
\bibitem{cdf-svx} D.~Amidei \textit{et al.}, Nucl. Instrum. Methods Phys. Res. A
{\bf 350}, 73 (1994); S.~Cihangir \textit{et al.}, Nucl. Instrum. Methods Phys. Res. A {\bf 360}, 137 (1995).
\bibitem{kane-gluino} G.~L.~Kane and S.~Mrenna, \prl{77}, 3502 (1996).
\end{thebibliography}
\end{document}